\documentclass[twocolumn,referee]{svjour3}
\usepackage{graphicx}
\usepackage{epstopdf}
\usepackage{amsmath}
\usepackage{amssymb}
\usepackage{float}
\usepackage{bm}

\input{epsf}
\begin{document}

\title{Adiabatic hyperspherical analysis of realistic nuclear potentials}

\author{K.~M. Daily \and Alejandro Kievsky \and Chris H. Greene}
\institute{
  K.~M. Daily \at
  \email{daily5@purdue.edu}
  Department of Physics and Astronomy,
  Purdue University,
  West Lafayette, Indiana 47907, USA \and
  Alejandro Kievsky \at
  Instituto Nazionale di Fisica Nucleare,
  Largo Pontecorvo 3, 56100 Pisa, Italy \and
  Chris H. Green \at
  Department of Physics and Astronomy,
  Purdue University,
  West Lafayette, Indiana 47907, USA   }

\date{\today}

\begin{abstract}
  Using the hyperspherical adiabatic method with
  the realistic nuclear potentials Argonne V14, Argonne V18,
  and Argonne V18 with the Urbana IX three-body potential,
  we calculate the adiabatic potentials and the triton bound state energies.
  We find that a discrete variable representation with the slow variable
  discretization method along the hyperradial degree of freedom results in
  energies consistent with the literature.
  However, using a Laguerre basis results in missing energy,
  even when extrapolated to an infinite number of basis functions
  and channels.
  We do not include the isospin $T=3/2$ contribution in our analysis.
\end{abstract}

\maketitle

\section{Introduction}
The hyperspherical adiabatic representation is well-established
in atomic physics~\cite{macek_properties_1968,lin_correlations_1974}.
Few-nucleon problems have also taken advantage of this technique
in solving simple model nuclear potentials~\cite{garrido_integral_2012}.
Though convergence with respect to the number of included channels
is typically favorable
in short-range low-energy atomic calculations~\cite{wang_hyperspherical_2012},
we set out to check the convergence in the three-body nuclear problem.
In particular,
we test whether a typical orthogonal basis is flexible enough to
handle the complicated nonadiabatic couplings.

%%%%%
The solution of the adiabatic Hamiltonian, eigenvalues (adiabatic potentials)
and eigenvectors (channel functions) enters in the construction of a set of
hyperradial equations that can be solved for bound or scattering states. This
gives to the hyperspherical adiabatic representation a great flexibility.
Most of the application of this technique has been done to determine bound state
solutions, however in recent years several applications to determine scattering
observables has been studied as well~\cite{garrido_breakup_2014}.
%%%%

In the present work, we use a hyperspherical harmonic basis to diagonalize 
the adiabatic Hamiltonian using
the Argonne V14 (AV14)~\cite{wiringa_nucleon-nucleon_1984},
Argonne V18 (AV18)~\cite{wiringa_accurate_1995},
and AV18 with the Urbana IX (UIX)~\cite{pieper_realistic_2001}
three-body nuclear potentials.
As a first step, to solve the adiabatic Hamiltonian
we use a basis of symmetrized hyperspherical harmonics.
In a second step,
we solve the set of coupled hyperradial equations to determine the three-nucleon
bound state using
either a Laguerre polynomial basis or discrete variable representation (DVR).
We compare the effectiveness of the two bases. This work can be seen as a first
step in the application of the DVR technique, currently used in atomic physics,
to the three-nucleon problem having in mind, as a final step, the treatment of
the continuum spectrum.

Section~\ref{sec_theory} describes the hyperspherical coordinate description
we use to solve the three-nucleon problem.
Section~\ref{sec_pcurves} shows the solutions to the adiabatic Hamiltonian.
Section~\ref{sec_energies} compares two methods in determining the
three-body bound state and Sec.~\ref{sec_conclusion} concludes.

\section{Theoretical background}
\label{sec_theory}
We concern ourselves only with the relative Hamiltonian $H_{\rm rel}$.
We recast $H_{\rm rel}$ in hyperspherical
coordinates~\cite{macek_properties_1968,lin_correlations_1974}
in terms of five hyperangles denoted by $\vec{\Omega}$
and a single length, the hyperradius $R$.
The relative Hamiltonian is then a sum of the 
hyperradial kinetic energy,
the hyperangular kinetic energy, and
the interaction potential,
\begin{align}
  \label{eq_SErel}
  H_{\rm rel} = - \frac{\hbar^2}{2 m}
                \left( \frac{1}{R^{5}}\frac{\partial}{\partial R}R^{5}
                \frac{\partial}{\partial R}
  -\frac{\vec{\Lambda}^2}{R^2}\right)
  + V_{\rm int}(R,\vec{\Omega}).
\end{align}
Here, $m$ is the average nucleon mass and $\hbar$ is Planck's constant.
The exact form
of the square of the grand angular momentum operator $\vec{\Lambda}^2$
depends on the choices of the Jacobi vectors and of the hyperangles
and is not needed here.

The solution $\Psi_E(R,\vec{\Omega})$ 
to Eq.~\eqref{eq_SErel} is expanded in terms of the
radial functions $R^{-5/2} F_{E\nu}(R)$ and
the channel functions $\Phi_{\nu}(R;\vec{\Omega})$,
\begin{align}
\label{eq_psi_expansion}
\Psi_E(R,\vec{\Omega}) = 
R^{-5/2}\sum_{\nu} F_{E\nu}(R)\Phi_{\nu}(R;\vec{\Omega}).
\end{align}
The channel functions at a fixed hyperradius $R$ form
a complete orthonormal set over the hyperangles,
\begin{align}
  \int \,d\vec{\Omega} \; \Phi^*_{\nu}(R;\vec{\Omega}) \Phi_{\nu'}(R;\vec{\Omega}) 
  = \delta_{\nu \nu'},
\end{align}
and are the solutions to the adiabatic Hamiltonian $H_{\rm ad}(R,\vec{\Omega})$,
\begin{align}
  H_{\rm ad}(R,\vec{\Omega}) \Phi_{\nu}(R;\vec{\Omega}) 
  = U_{\nu}(R) \Phi_{\nu}(R;\vec{\Omega}),
\end{align}
where
\begin{align}
  \label{eq_Had}
  H_{\rm ad}  = & 
  \frac{\hbar^2}{2m} \frac{ \vec{\Lambda}^2 + 15/4 }{R^2} 
  + V_{\rm int}(R,\vec{\Omega}).
\end{align}

After applying Eq.~\eqref{eq_SErel} on the expansion 
Eq.~\eqref{eq_psi_expansion}
and projecting from the left onto the channel functions,
the Schr\"odinger equation reads
\begin{align}
\label{eq_SE_W}
&
\left( 
  -\frac{\hbar^2}{2m}\frac{d^2}{d R^2} 
  + U_{\nu}(R)  -E
\right) F_{E \nu}(R)
\\ \nonumber & \qquad 
-\frac{\hbar^2}{2m} \sum_{\nu'}
  \left( 2 P_{\nu \nu'}(R)\frac{d}{d R} + Q_{\nu \nu'}(R) \right)
  F_{E \nu'}(R)
= 0.
\end{align}
The hyperspherical Schr\"odinger equation Eq.~\eqref{eq_SE_W}
is solved in a two step procedure.
First, $H_{\rm ad}(R,\vec{\Omega})$ is solved parametrically in $R$ for
the adiabatic potential curves $U_{\nu}(R)$.
In a second step,
the coupled set of one-dimensional equations in $R$ are solved. 
In Eq.~\eqref{eq_SE_W}, $P_{\nu\nu'}$ and $Q_{\nu\nu'}$ 
represent the coupling between channels,
where 
\begin{align}
\label{eq_Pcoupling}
  P_{\nu \nu'}(R) = \bigg\langle \Phi_{\nu} \bigg| 
  \frac{\partial \Phi_{\nu'}}{\partial R} \bigg\rangle_{\vec{\Omega}}
\end{align}
and
\begin{align}
\label{eq_Qcoupling}
  Q_{\nu \nu'}(R) = \bigg\langle \Phi_{\nu} \bigg| 
  \frac{\partial^2 \Phi_{\nu'}}{\partial R^2} \bigg\rangle_{\vec{\Omega}}.
\end{align}
The brackets indicate that the integrals are taken only over the
hyperangle $\vec{\Omega}$ with the hyperradius $R$ held fixed.

\section{Adiabatic potential curves}
\label{sec_pcurves}
We expand the channel functions, $\Phi_{\nu}(R;\vec{\Omega})$, 
using a basis of hyperspherical harmonics $\phi_{KM}(\vec{\Omega})$.
Here,
$K$ labels the grand angular momentum quantum number and
$M$ labels the different degenerate states for a fixed $K$.
The hyperspherical harmonics diagonalize
the hyperangular kinetic energy operator,
where $\vec{\Lambda}^2 \phi_{KM}(\vec{\Omega}) = K (K+4) \phi_{KM}(\vec{\Omega})$.
The main challenge comes from calculating the potential matrix elements
$\langle \phi_{KM}(\vec{\Omega})|V_{\rm int} | \phi_{KM}(\vec{\Omega}) \rangle$
at a fixed hyperradius,
which has been worked out by the second author using the technique of
Ref.~\cite{kievsky_high_1997}.

The hyperspherical harmonics are chosen to have certain symmetry properties.
For example,
subsets of basis functions are chosen to have orbital angular momenta $L_{12}$,
$L_{12,3}$, and $L$.
These correspond to the orbital angular momentum along the first Jacobi vector,
the second Jacobi vector, and the total orbital angular momenta, respectively.
Additionally,
the spin $S_{12}$ and isospin $T_{12}$ along the first Jacobi vector,
and the total spin $S$ and isospin $T$ are fixed.
Other restrictions include that $L_{12}+S_{12}+T_{12}$
must be odd for antisymmetrization
and $L$ and $S$ couple to $J=1/2$, the total angular momentum of the triton.
In general,
the sub-orbital angular momenta and spin are not good quantum numbers.
Thus,
the basis is not orthogonal and the adiabatic Hamiltonian, Eq.~\eqref{eq_Had},
is solved via a generalized eigenvalue problem at each hyperradius.

Table~\ref{tab_1} shows an example of a set of basis functions.
\setlength{\tabcolsep}{4pt}
\begin{table}
  \begin{tabular}{c|*{9}{c}}
    Channel & $L_{12}$ & $L_{12,3}$ & $L$ & $S_{12}$ & $T_{12}$ & $S$ & $T$ & $K_{min}$ & $K_{\rm max}$ \\ \hline
    1 &   0 &  0 &  0 &  1 &  0 &  1/2 &  1/2 &  0  & 120 \\
    2 &   0 &  0 &  0 &  0 &  1 &  1/2 &  1/2 &  4  & 120 \\
    3 &   2 &  0 &  2 &  1 &  0 &  3/2 &  1/2 &  2  & 120 \\
    4 &   0 &  2 &  2 &  1 &  0 &  3/2 &  1/2 &  4  & 90 \\
    5 &   2 &  2 &  0 &  1 &  0 &  1/2 &  1/2 &  10 & 90 \\
    6 &   2 &  2 &  2 &  1 &  0 &  3/2 &  1/2 &  6  & 90 \\
    7 &   2 &  2 &  1 &  1 &  0 &  1/2 &  1/2 &  6  & 90 \\
    8 &   2 &  2 &  1 &  1 &  0 &  3/2 &  1/2 &  4  & 90 \\
    9 &   1 &  1 &  0 &  1 &  1 &  1/2 &  1/2 &  6  & 60 \\
    10&   1 &  1 &  1 &  1 &  1 &  1/2 &  1/2 &  2  & 60 \\
    11&   1 &  1 &  1 &  1 &  1 &  3/2 &  1/2 &  10 & 60 \\
    12&   1 &  1 &  2 &  1 &  1 &  3/2 &  1/2 &  8  & 60 \\
    13&   1 &  1 &  0 &  0 &  0 &  1/2 &  1/2 &  12 & 60 \\
    14&   1 &  1 &  1 &  0 &  0 &  1/2 &  1/2 &  8  & 60 \\
    15&   2 &  2 &  0 &  0 &  1 &  1/2 &  1/2 &  16 & 40 \\
    16&   2 &  2 &  1 &  0 &  1 &  1/2 &  1/2 &  12 & 40 \\
    17&   1 &  3 &  2 &  1 &  1 &  3/2 &  1/2 &  10 & 40 \\
    18&   3 &  1 &  2 &  1 &  1 &  3/2 &  1/2 &  12 & 40
  \end{tabular}
  %% \caption{Example of functions used in the hyperangular basis.
  %%   $L_{12}$, $S_{12}$ and $T_{12}$ indicate the orbital angular momentum, spin, and isospin
  %%   of the (12) pair,
  %%   while $L$, $S$ and $T$ are the total orbital angular momentum, spin, and isospin.
  %%   $L_{12,3}$ is the orbital angular momentum of the third particle relative to the (12) pair.
  %%   $K_{min}$ ($K_{\rm max}$) is the minimum (maximum) grand hyperangular quantum number.
  %%   $K_{\rm max}$ is adjusted for better convergence; this example was used to generate data
  %%   in Fig.~\ref{???}.}
  \caption{Channel quantum numbers used in the hyperangular basis.
    $L_{12}$, $L_{12,3}$, and $L$, indicate the orbital angular momentum
    of the (12) pair,
    that of the third particle relative to the (12) pair, and the total.
    $S_{12}$ and $T_{12}$ indicate the spin and isospin of the (12) pair,
    while $S$ and $T$ are the total spin and isospin.
    $K_{min}$ ($K_{\rm max}$) is the minimum (maximum)
    grand hyperangular quantum number.
    The given values are used to generate data
    in Figs.~\ref{fig_pot}(a), \ref{fig_Lag} and~\ref{fig_DVR}.}
\label{tab_1}
\end{table}
The different quantum numbers are listed in each column with a given set of channels
labeled 1 through 18.
The number of basis functions comes
from the range of grand angular quantum numbers $K$ for each channel,
where $K$ varies from $K_{min}$, to ensure linear independence, 
to $K_{\rm max}$ in steps of two to preserve parity.
The total number of basis functions represented here is 617,
where 227 have $L=0$, 185 have $L=1$, and 205 have $L=2$. This particular selection
of channels and values of $K_{\rm max}$ describes the three-nucleon bound state
with an accuracy of about $1$ keV~\cite{kievsky_high_1997}.

\begin{figure}
  \centering
  %\vspace*{+1.5cm}
  \includegraphics[angle=0,width=0.4\textwidth]{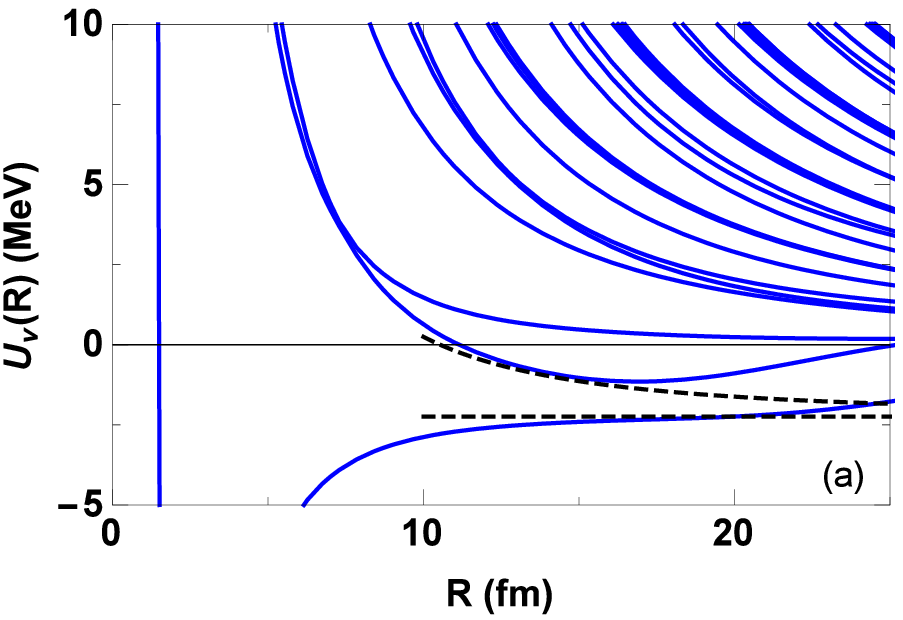} 
  \includegraphics[angle=0,width=0.4\textwidth]{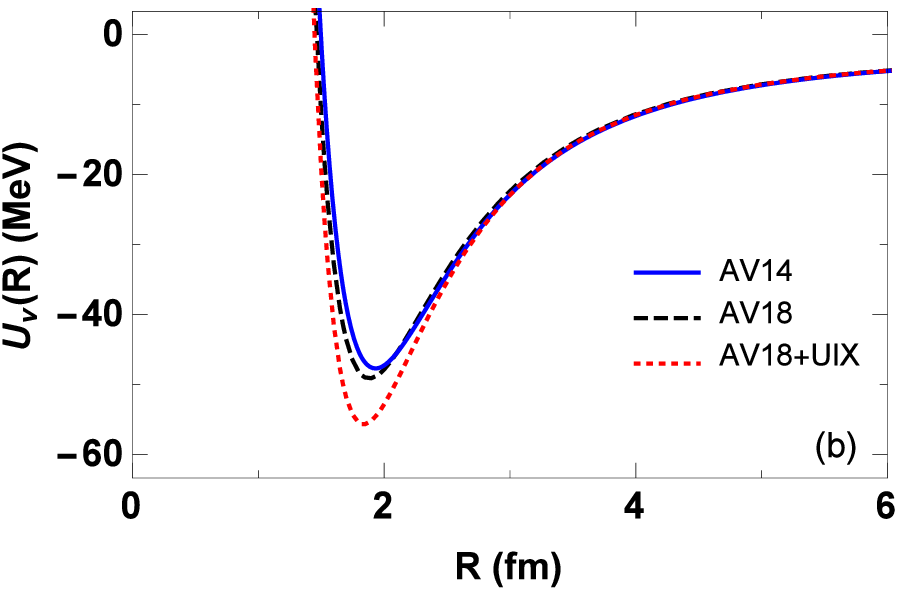} 
  %\vspace*{0.1cm}
  \caption{(Color online) 
    Adiabatic potential curves.
    Panel (a) shows the adiabatic potential curves for the AV14 potential
    using the basis set in Table~\ref{tab_1}.
    The dashed lines show the known asymptotic limits for the two lowest
    potential curves, approaching the deuteron binding energy of -2.22MeV.
    Panel (b) shows the ground state adiabatic potential curves for
    three different nuclear potentials.
  }
  \label{fig_pot}
\end{figure}

The solid lines in Figure~\ref{fig_pot}(a)
are the adiabatic potential curves
generated from the basis in Table~\ref{tab_1} for the AV14 nuclear potential.
The full ground state potential curve can be seen in panel (b).
There are a number of lines that asymptotically approach the zero energy
threshold,
indicating that these curves represent the fragmentation
into three free particles.
Two lines approach a negative energy threshold of -2.22MeV,
such that these channels indicate the fragmentation into a deuteron
and a free neutron.
The two lowest potential curves are not converged,
but the known asymptotic limits are indicated by dashed lines.
In practice,
the potentials are smoothly connected to their known asymptotic behavior.
The lowest potential corresponds to an $s$-wave configuration between
the deuteron and the neutron, where there is no angular momentum barrier.
The second lowest potential corresponds to a $d$-wave configuration,
where there is a significant angular momentum barrier.

It is known that the hyperspherical harmonic basis works well at
small hyperradius,
but requires a large number of basis functions to reach similar convergence
at large hyperradius.
We have seen that with the inclusion of more basis functions,
e.g. going to 1052 basis functions
with 362 at $L=0$, 335 at $L=1$, and 355 at $L=2$,
that the lowest two potential curves in Fig.~\ref{fig_pot}(a)
are indistinguishable from their asymptotic behavior on the scale shown.
Specifically,
$K_{\rm max}$ is taken to be
150 for channels 1 to 6,
140 for channels 7 and 8,
110 for channels 9 to 14,
and 90 for channels 15 to 18.
%This larger basis is used throughout the rest of this work.
Figure~\ref{fig_pot}(b) uses this larger basis to
show the lowest adiabatic potential curve
for three different nuclear potentials.
Solid, dashed, and dotted lines are for the AV14, AV18, and AV18 with UIX
potentials, respectively.
The difference between the different nuclear potentials is most clearly
observed in the lowest adiabatic potential.
The inclusion of the three-body potential, e.g.,
produces the lowest potential curve.
On the scale shown in Fig.~\ref{fig_pot}(a),
there is no visible difference between the channels of the excited states.
The next section describes the calculation of the bound state supported
by the lowest adiabatic potentials.

\section{Bound state energies}
\label{sec_energies}
With the adiabatic potentials $U_{\nu}(R)$,
we solve the hyperradial Schr\"odinger equation Eq.~\eqref{eq_SE_W}
to determine the bound state energy of the triton.
As a first approach,
we use a seven-point finite difference method to determine the $P$ and $Q$
matrix elements,
together with a basis of Laguerre polynomials in the hyperradius $R$.
There are two convergence criteria to consider.
First,
sufficiently many Laguerre polynomials must be used to describe the
hyperradial wave functions.
Second,
sufficiently many channel functions must be included to reach convergence.

\begin{figure}
  \centering
  %\vspace*{+1.5cm}
  \includegraphics[angle=0,width=0.4\textwidth]{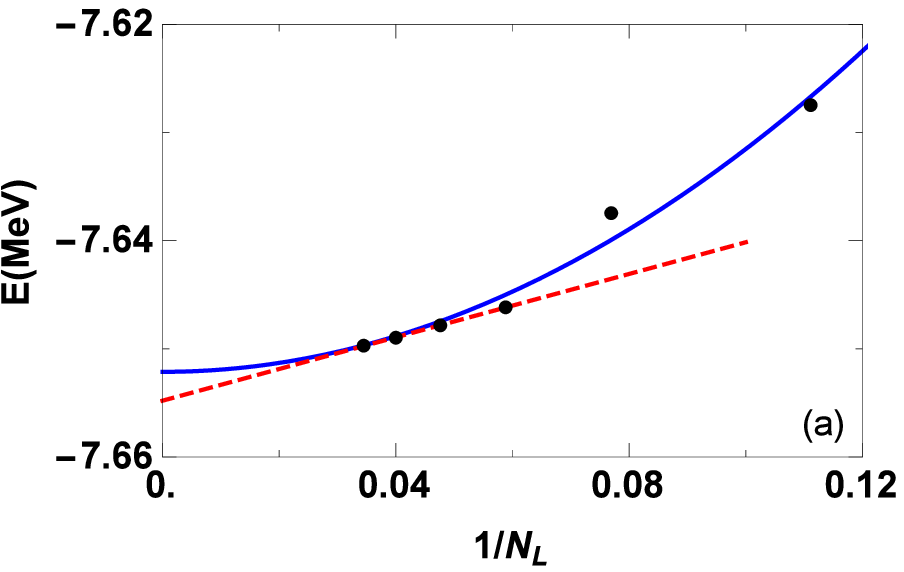} 
  \includegraphics[angle=0,width=0.4\textwidth]{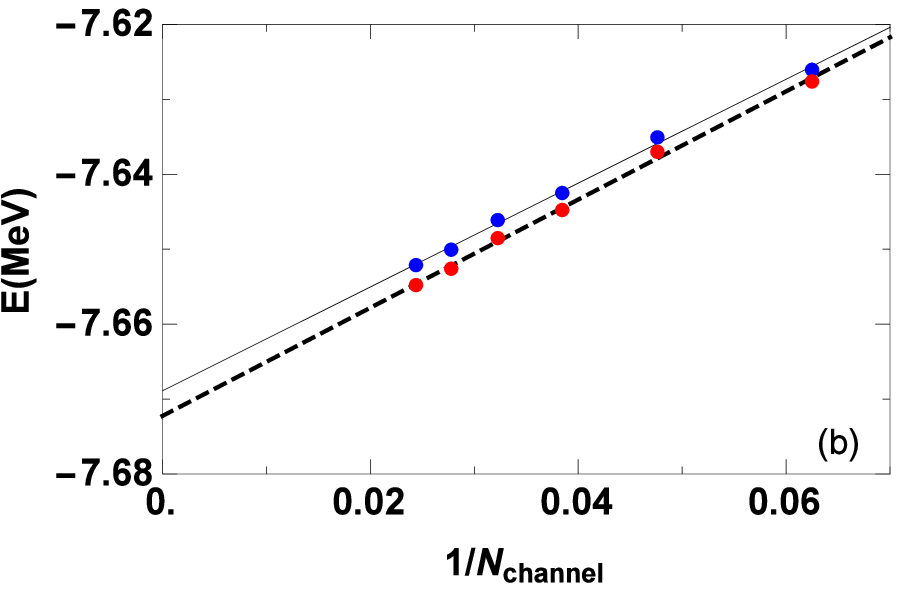} 
  %\vspace*{0.1cm}
  \caption{(Color online) 
    Extrapolations used to estimate the triton energy for the AV14 potential
    using a Laguerre basis in the hyperradius.
    The upper panel shows the change in energy
    as a function of the number of Laguerre polynomials $N_L$ for 41 channel functions.
    The solid lines show strictly quadratic and linear fits to subsets of the data points. 
    The lower panel shows the extrapolated values for different numbers of channel functions 
$N_{\rm channel}$.
    The upper and lower points (dashed and solid lines) correspond to the quadratic
    and linear fits, respectively.
    The lines show linear extrapolations to the infinite channel limit.
  }
  \label{fig_Lag}
\end{figure}

Figure~\ref{fig_Lag}(a)
shows a convergence study for the AV14 potential
as a function of the number of Laguerre basis functions $N_L$
with the lowest 41 channel functions included.
The hyperangular basis used is that described in Table~\ref{tab_1}.
The solid (dashed) line is a strictly quadratic (linear) fit to
the data from $1/N_L=0.03$ to 0.12 (0.03 to 0.06).
The two fits are used to estimate the energy assuming an infinite
set of Laguerre basis functions.
These extrapolations are calculated at different numbers of included channels
$N_{\rm channel}$ and shown in Fig.~\ref{fig_Lag}(b).
The upper (lower) points are from the quadratic (linear) extrapolations.
%like those shown in Fig.~\ref{fig_Lag}(a).
Solid and dashed lines are linear fits
to the data points of Fig.~\ref{fig_Lag}(b).
Extrapolating to an infinite number of channels gives an estimate of
$-7.671(2)$MeV for the triton binding energy.

This technique seems to miss about 10keV of binding energy when compared
with other estimates~\cite{perez_triton_2014},
or worse if estimating the energy by the last data point
and not extrapolating to an infinite basis.
We suggest that the error comes from the finite-difference method
to calculate the $P$ and $Q$ matrix elements.
The couplings show complicated behavior at small $R$ that may not be
captured by this technique.
We propose using a discrete variable
representation~\cite{light_generalized_1985,rescigno_numerical_2000} in $R$
together with the slow variable discretization method (SVD)~\cite{tolstikhin_slow_1996}.
Thus,
the $P$ and $Q$ matrices are not needed since the hyperradial solution 
is solved in sectors that are exactly diagonalized.
The only additional requirement is that the phase of each diagonalized
sector must match between sector boundaries.

\begin{figure}
  \centering
  %\vspace*{+1.5cm}
  \includegraphics[angle=0,width=0.4\textwidth]{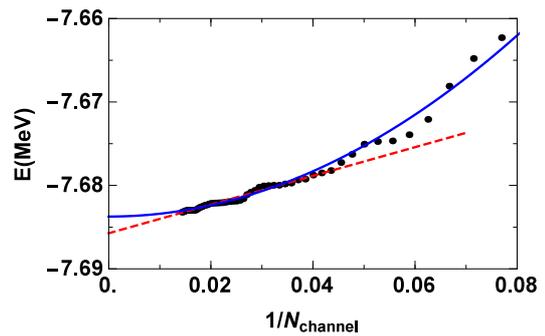} 
  %\vspace*{0.1cm}
  \caption{(Color online) 
    Extrapolations used to estimate the triton energy for the AV14 potential
    using a DVR basis in the hyperradius with the SVD method.
    Solid and dashed lines show strictly quadratic and linear fits to the data points
    from $1/N_{\rm channel}$=[0,0.04].
  }\label{fig_DVR}
\end{figure}

Figure~\ref{fig_DVR}
shows the energy
as a function of the number of included channels $N_{\rm channel}$.
For this data,
we use a total of 40 SVD sectors,
with 20 ranging from $R=0.2$fm to 3fm
and 20 ranging from $R=3$fm to 50fm.
Each SVD sector consists of a DVR basis of five points.
The hyperangular basis used is that described in Table~\ref{tab_1}.
We find that the data changes in the fifth significant digit when
we increase the number of DVR points to 10.
Further,
we find that the results are robust given changes in the number of sectors,
so long as the potential minima is covered by enough points
and the maximum range extends to the asymptotic limit at about 20fm.
The solid (dashed) line of Fig.~\ref{fig_DVR} is a strictly
quadratic (linear) fit to the data
including all points where $1/N_{\rm channel}<0.04$.

Compared to the Laguerre basis,
we achieve a noticeably lower energy for the same number of included channels.
Said differently,
if we do not extrapolate to an infinite number of channels,
the DVR energies would produce a better estimate than the Laguerre basis.
The DVR basis is not a variational method, however,
thus we include more channels in our DVR analysis
(the left-most data point includes the 70 lowest channels).
Regardless,
the method shows a clear convergence
and we estimate the triton binding energy to be $-7.685(1)$MeV.

The analysis is identical for the other nuclear potentials,
with Fig.~\ref{fig_DVR} being qualitatively similar for each.
Using the larger hyperspherical basis described at the end of
Sec.~\ref{sec_pcurves},
we estimate the triton binding energy to be $-7.685(1)$MeV,
$-7.619(2)$MeV, and $-8.475(2)$MeV
for the AV14, AV18, and AV18 with UIX potentials, respectively.
These results are in complete agreement with those given in the benchmark
of Ref.~\cite{nogga_three-nucleon_2003} showing the great flexibility of the
SVD technique to treat the complexity of the adiabatic couplings.

\section{Conclusion and Outlook}
\label{sec_conclusion}
In this paper we calculated the adiabatic potentials curves
for the three nucleon problem using realistic nuclear potentials
in order to calculate the three-nucleon bound state.
In calculating the triton binding energy,
we find that the SVD method with a DVR basis gives results consistent
with the literature~\cite{nogga_three-nucleon_2003},
while using a naive finite-difference method leads to inconsistent results.
This suggests that the channel couplings are handled well by
the DVR basis.

Though the binding energy of three nucleons can be calculated by
different techniques (see Ref.~\cite{nogga_three-nucleon_2003} and
references therein),
this method may prove useful in calculating scattering data.
As a preliminary study,
we attempted to propagate the $R$ matrix in the SVD method,
but the hyperspherical harmonic basis used does not produce convergence at large values
of the hyperradius.
This led to difficulties as 
we could not accurately connect the potentials and coupling matrix elements
to their asymptotic behavior at large hyperradius
in order to switch to a traditional $R$ matrix propagation method.
To get around this limitation for scattering states below the three-nucleon
breakup,
we could use the method described in Ref.~\cite{kievsky_theoretical_2012},
which only requires information from the bound state calculations
in order to estimate the $S$ matrix elements. For determining the breakup
amplitude a larger basis should be necessary in conjunction with the method
described in Ref.~\cite{garrido_breakup_2014}.

\begin{acknowledgements}
Support by the US Deptartment of Energy, Office of Science through Grant No. 
DE-SC0010545 is gratefully acknowledged.
\end{acknowledgements}


\begin{thebibliography}{10}
\providecommand{\url}[1]{{#1}}
\providecommand{\urlprefix}{URL }
\expandafter\ifx\csname urlstyle\endcsname\relax
  \providecommand{\doi}[1]{DOI \discretionary{}{}{}#1}\else
  \providecommand{\doi}{DOI \discretionary{}{}{}\begingroup
  \urlstyle{rm}\Url}\fi

\bibitem{macek_properties_1968}
J.~Macek, J. Phys. B: At. Mol. Opt. \textbf{1}(5), 831 (1968).
\newblock \urlprefix\url{http://stacks.iop.org/0022-3700/1/i=5/a=309}

\bibitem{lin_correlations_1974}
C.D. Lin, Physical Review A \textbf{10}(6), 1986 (1974).
\newblock \doi{10.1103/PhysRevA.10.1986}.
\newblock \urlprefix\url{http://link.aps.org/doi/10.1103/PhysRevA.10.1986}

\bibitem{garrido_integral_2012}
E.~Garrido, C.~Romero-Redondo, A.~Kievsky, M.~Viviani, Physical Review A
  \textbf{86}(5), 052709 (2012).
\newblock
  \urlprefix\url{http://journals.aps.org/pra/abstract/10.1103/PhysRevA.86.052709}

\bibitem{wang_hyperspherical_2012}
J.~Wang, Hyperspherical approach to quantal three-body theory.
\newblock Ph.D. thesis, University of Colorado, Boulder (2012)

\bibitem{garrido_breakup_2014}
E.~Garrido, A.~Kievsky, M.~Viviani, Physical Review C \textbf{90}, 014607
  (2014).
\newblock \doi{10.1103/PhysRevC.90.014607}.
\newblock \urlprefix\url{http://link.aps.org/doi/10.1103/PhysRevC.90.014607}

\bibitem{wiringa_nucleon-nucleon_1984}
R.B. Wiringa, R.A. Smith, T.L. Ainsworth, Phys. Rev. C \textbf{29}(4), 1207
  (1984).
\newblock \doi{10.1103/PhysRevC.29.1207}.
\newblock \urlprefix\url{http://link.aps.org/doi/10.1103/PhysRevC.29.1207}

\bibitem{wiringa_accurate_1995}
R.B. Wiringa, V.G.J. Stoks, R.~Schiavilla, Physical Review C \textbf{51}(1), 38
  (1995).
\newblock
  \urlprefix\url{http://journals.aps.org/prc/abstract/10.1103/PhysRevC.51.38}

\bibitem{pieper_realistic_2001}
S.~Pieper, Phys. Rev. C \textbf{64}(1) (2001).
\newblock \doi{10.1103/PhysRevC.64.014001}

\bibitem{kievsky_high_1997}
A.~Kievsky, L.E. Marcucci, S.~Rosati, M.~Viviani, Few-Body Systems
  \textbf{22}(1), 1 (1997).
\newblock \doi{10.1007/s006010050049}.
\newblock \urlprefix\url{http://dx.doi.org/10.1007/s006010050049}

\bibitem{perez_triton_2014}
R.N. Pérez, E.~Garrido, J.E. Amaro, E.~Ruiz~Arriola, Phys. Rev. C
  \textbf{90}(4), 047001 (2014).
\newblock \doi{10.1103/PhysRevC.90.047001}.
\newblock \urlprefix\url{http://link.aps.org/doi/10.1103/PhysRevC.90.047001}

\bibitem{light_generalized_1985}
J.C. Light, I.P. Hamilton, J.V. Lill, The Journal of Chemical Physics
  \textbf{82}(3), 1400 (1985).
\newblock \doi{10.1063/1.448462}.
\newblock
  \urlprefix\url{http://scitation.aip.org/content/aip/journal/jcp/82/3/10.1063/1.448462}

\bibitem{rescigno_numerical_2000}
T.N. Rescigno, C.W. McCurdy, Physical review A \textbf{62}(3), 032706 (2000).
\newblock
  \urlprefix\url{http://journals.aps.org/pra/abstract/10.1103/PhysRevA.62.032706}

\bibitem{tolstikhin_slow_1996}
O.I. Tolstikhin, S.~Watanabe, M.~Matsuzawa, Journal of Physics B: Atomic,
  Molecular and Optical Physics \textbf{29}(11), L389 (1996).
\newblock \urlprefix\url{http://iopscience.iop.org/0953-4075/29/11/001}

\bibitem{nogga_three-nucleon_2003}
A.~Nogga, A.~Kievsky, H.~Kamada, W.~Glöckle, L.~Marcucci, S.~Rosati,
  M.~Viviani, Physical Review C \textbf{67}(3) (2003).
\newblock \doi{10.1103/PhysRevC.67.034004}.
\newblock \urlprefix\url{http://link.aps.org/doi/10.1103/PhysRevC.67.034004}

\bibitem{kievsky_theoretical_2012}
A.~Kievsky, M.~Viviani, L.E. Marcucci, Physical Review C \textbf{85}(1) (2012).
\newblock \doi{10.1103/PhysRevC.85.014001}.
\newblock \urlprefix\url{http://link.aps.org/doi/10.1103/PhysRevC.85.014001}

\end{thebibliography}
\end{document}